\documentclass[twocolumn,amsmath,amssymb,pre,showpacs]{revtex4}

\usepackage{graphicx}
\usepackage{dcolumn}
\usepackage{bm}
\usepackage{amssymb}
\usepackage{amsmath}
\usepackage{amsthm,amsfonts}
\DeclareMathOperator{\csch}{csch}

\begin{document}
\bibliographystyle{aip}

\title{Heat Rectification on the XX chain}

\author{ Saulo H. S. Silva$^{1}$, Gabriel T. Landi$^{2}$, Raphael C. Drumond$^{3}$ and Emmanuel Pereira$^{1}$}
 \email{emmanuel@fisica.ufmg.br}
\affiliation{$^{1}$Departamento de F\'{\i}sica, Instituto de Ci\^encias Exatas, Universidade Federal de Minas Gerais, 30123-970, Belo Horizonte, Minas Gerais, Brazil\\
$^{2}$Instituto de Física da Universidade de São Paulo, 05314-970 São Paulo, Brazil\\
$^{3}$Departamento de Matemática, Instituto de Ciências Exatas,
Universidade Federal de Minas Gerais, 30123-970, Belo Horizonte, Minas Gerais, Brazil}

\begin{abstract}
In order to better understand the minimal ingredients for thermal rectification, we perform a detailed investigation of a 
simple spin chain, namely, the open $XX$ model with a Lindblad dynamics involving global dissipators. We use a Jordan-Wigner transformation to derive a mathematical formalism to compute the heat currents and other properties of
the steady state. 
 We have rigorous results to prove
the occurrence of thermal rectification even for slightly asymmetrical chains. Interestingly, we describe cases where the rectification does not 
decay to zero as we increase the system size, that is, the rectification
remains finite in the thermodynamic limit. We also describe some numerical
results for more asymmetrical chains. The presence of thermal rectification
in this simple model indicates that the phenomenon is of general occurrence
in quantum spin systems.  
\end{abstract}


\def \Z {\mathbb{Z}}
\def \R {\mathbb{R}}
\def \La {\Lambda}
\def \la {\lambda}
\def \ck {l}
\def \F {\mathcal{F}}
\def \M {\mathcal{M}}
\newcommand {\md} [1] {\mid\!#1\!\mid}
\newcommand {\be} {\begin{equation}}
\newcommand {\ee} {\end{equation}}
\newcommand {\ben} {\begin{equation*}}
\newcommand {\een} {\end{equation*}}
\newcommand {\bg} {\begin{gather}}
\newcommand {\eg} {\end{gather}}
\newcommand {\ba} {\begin{align}}
\newcommand {\ea} {\end{align}}
\newcommand {\tit} [1] {``#1''}


\maketitle

\let\a=\alpha \let\b=\beta \let\d=\delta \let\e=\varepsilon
\let\f=\varphi \let\g=\gamma \let\h=\eta    \let\k=\kappa \let\l=\lambda
\let\m=\mu \let\n=\nu \let\o=\omega    \let\p=\pi \let\ph=\varphi
\let\r=\rho \let\s=\sigma \let\t=\tau \let\th=\vartheta
\let\y=\upsilon \let\x=\xi \let\z=\zeta
\let\D=\delta \let\G=\Gamma \let\L=\Lambda \let\Th=\Theta
\let\P=\Pi \let\Ps=\Psi \let\Si=\Sigma \let\X=\Xi
\let\Y=\Upsilon

\section{Introduction}
One of the fundamental issues of nonequilibrium statistical physics is the
derivation of transport laws from the underlying microscopic dynamics. In particular, a theme of general interest is the investigation of energy
transport, which involves two main mechanisms, the conduction by electricity and by heat, issues, however, with quite different status in the
literature. On the one hand, the success of modern electronics since the invention of the transistor is well known, with huge repercussion in our daily
lives. On the other hand, we see a slow progress of phononics, the counterpart of electronics dedicated to the study and manipulation of heat current. Heat analogs of electronic devices, such as transistors and 
gates have been already proposed \cite{BLiRMP}, but the absence of a 
feasible and efficient thermal diode, the basic ingredient of these devices, makes difficult a considerable advance. Thermal diode or thermal rectifier is a device in which heat has a preferable direction to flow, more precisely, the magnitude of the heat current changes as we invert the 
device between two thermal baths. And so, the first obvious ingredient for
the occurrence of rectification is the existence of an asymmetry in the system.

The most usual models for the study of heat conduction in insulating solids
is given, since Debye \cite{Deb} and Peierls \cite{Pei}, by chains of classical harmonic or anharmonic oscillators. Unfortunately, in the more
treatable harmonic version there is no thermal rectification. Even for the
harmonic classical system with inner self-consistent stochastic reservoirs
\cite{BLL}, it is proved the absence of thermal rectification \cite{PLA}.
It is intersting to recall that such system obeys the Fourier law, that
does not hold in purely harmonic chains \cite{RLL}, showing that the inner
reservoirs indeed represent some vestiges of anharmonicity, which, however,
are not enough for the occurrence of thermal rectification. 

The search for the minimal ingredients sufficient to guarantee rectification is a fundamental and difficult problem in transport theory. In this 
direction we recall the study of simple models, avoiding intricate details
which may hide the ingredients. For example, we recall the establishment of rectification in Ref.\cite{WPC}, a toy model of alternating graded bars and bullets. There, one learns that the existence of a local temperature dependent thermal conductivity together with the graded structure assure the rectification. 

Besides this recurrent study of classical oscillators and related models, it is important to stress the present increasing interest in the study of
energy transport at the quantum scale, motivated, e.g., by the emerging field 
of quantum thermodynamics and the advances allowing the manipulation of quantum systems. In particular, there are recurrent 
investigations of quantum spin models, which involve problems in connection with different areas: condensed matter, cold atoms, quantum information, etc. 

In this direction, rectification in the boundary driven $XXZ$ spin $1/2$ model (with polarization at the edges) is shown in Ref.\cite{GL1},
for the spin current in the case of a homogeneous chain with asymmetrical
external magnetic field, and it is shown in Ref.\cite{SPL} for the energy
current in a graded chain. We recall that the $XXZ$ chains are the archetypal models for open quantum spin systems. Interestingly, in Ref.\cite{GL1} it is shown the absence of spin rectification in the system
with zero anisotropy  parameter $\Delta$ (coefficient of $\sigma_{j}^{z}\sigma_{j+1}^{z}$). For $\Delta \neq 0$, rectification is observed. As the $XXZ$ model can be mapped into a problem of bosons with
creation and annihilation operators, with quadratic terms and a quartic one
proportional to $\Delta$ (Tonks-Girardeau model), the vanishing of rectification in the absence of the quartic term is compared to the case
of classical oscillators, where there is no rectification in the absence of
anharmonicity (given by terms of order four or up in the potential).

Anyway, heat rectification has been described in some quadratic models with
proper arrangements, for example, the quantum Ising model is shown to rectify \cite{P1} if the intersite interaction is long enough to link the
first site (connected to the left bath) to the last one (connected to the
right bath). Otherwise, there is no rectification in such model.

In the present work, searching for simple quantum models showing rectification, that is, aiming to shed some light in the question of
minimal ingredients necessary for the occurrence of heat rectification, we
perform an analytical detailed investigation of the $XX$ spin $1/2$ model 
with some specific dissipators  and nearest neighbor interactions only. Even for a
slight asymmetric chain, we prove the occurrence of thermal rectification by
performing analytical computations. Interestingly, we describe cases of
heat rectification which does not decay to zero as the system size increases, that is, it remains finite in the thermodynamic limit. 
 We still show the rectification for more asymmetrical chains by using numerical techniques. The presence of 
heat rectification in this simple quadratic quantum spin model, i.e., in a simple system without intricate interactions, indicates that it is a ubiquitous phenomenon in the quantum context: for the occurrence of thermal rectification, it seems that we need only asymmetry in the system  and a thermal conductivity (or inner parameters) depending on temperature, and so, parameters which change as we invert the baths leading to rectification.

The rest of the paper is organized as follows. In section 2, we introduce
the model, the Jordan-Wigner transformation and some initial results. In
section 3, we describe the currents and some properties. In section 4, 
analytical results for the heat rectification are shown. Section 5 presents
some numerical results and section 6 is devoted to concluding remarks.

\section{Model and Preliminary Details}
Here we consider a one-dimensional quantum $XX$ spin chain with $N$ sites, described by the Hamiltonian 
\begin{equation}\label{eq:1}
    H=\sum_{j=1}^{N}\frac{h_{j}}{2}\sigma_{j}^{z}+\frac{1}{2}\sum_{j=1}^{N-1}\alpha_{j}(\sigma_{j}^{x}\sigma_{j+1}^{x}+\sigma_{j}^{y}\sigma_{j+1}^{y})\quad,
\end{equation}
\noindent
where the $\sigma_{j}^{i}$ are the usual Pauli matrices, $h_{j}$ is the external magnetic field acting on site $j$ and $\alpha_{j}$ is the exchange interaction between spins $j$ and $j+1$. The rectification will be directly associated with the asymmetry of the coefficients $h_{j}$ and $\alpha_{j}$ with respect to the left-right reflection of the chain.

These spin chains are coupled on the first and last sites to thermal reservoirs, kept at temperatures $T_{L}$ and $T_{R}$, respectively. They are modeled by an infinite number of bosonic degrees of freedom given by the Hamiltonian

\begin{equation}\label{eq:2}
    H_{B}^{i}=\sum_{l}\Omega_{i,l}a_{i,l}^{\dagger}a_{i,l}\quad,
\end{equation}
\noindent
where $a_{i,l}$ are a set of independent bosonic operators and $\Omega_{i,l}$ are the corresponding frequencies, which we assume to take on a quasi-continuum of values in the interval $[0,\infty)$. Moreover, the interaction with the first and last sites are assumed to take the form

\begin{equation}\label{eq:3}
\begin{aligned}
    H_{I}^{L}&=\sigma_{1}^{x}\sum_{i}g_{i}(a_{L,i}^{\dagger}+a_{L,i})\\
     H_{I}^{R}&=\sigma_{N}^{x}\sum_{i}g_{i}(a_{R,i}^{\dagger}+a_{R,i})\quad.
    \end{aligned}
\end{equation}

In order to proceed with the study, we recast the problem as a Lindblad master equation in the \textit{weak coupling regime}\cite{breuer}, describing the time evolution of the system's density matrix $\rho$ by

\begin{equation}\label{eq:4}
    \frac{d\rho}{dt}=-i[H,\rho]+\mathcal{D}_{L}+\mathcal{D}_{R}\quad,
\end{equation}
\noindent
where $\mathcal{D}_{L}$ and $\mathcal{D}_{R}$ are the Lindblad dissipators associated to the baths. It is possible to derive them from Eq.\eqref{eq:3} using the method of eigenoperators \cite{breuer}. 

Consider first only a single system-bath interaction, with Hamiltonian $H_{I}=A\otimes B$, where $A$ and $B$ are Hamiltonian operators of the system and the bath, respectively. We define 

\begin{equation}\label{eq:5}
\begin{aligned}
    \Gamma(\omega)&=\int_{-\infty}^{\infty}e^{i\omega t}\left<B(t)B(0)\right>dt\\
    &=\int_{-\infty}^{\infty}e^{i\omega t}tr\left\{e^{iH_{B} t}Be^{-iH_{B} t}B\frac{e^{-\frac{H_{B}}{T}}}{Z}\right\}dt\quad,
    \end{aligned}
\end{equation}
\noindent
that is, the Fourier transform of the bath correlations, evaluated for a bath thermal state, with temperature $T$ and partition function $Z=tr(e^{-\frac{H_{B}}{T}})$.

Let us define $\epsilon$ to be the eigenenergies of $H$ and $\Pi_{\epsilon}$ the corresponding projection operators onto the subspace corresponding to  $\epsilon$. From the \textit{weak coupling limit} derivation \cite{breuer} we define the eigenoperator corresponding to the bath-coupling $A$ as 

\begin{equation}\label{eq:6}
    A(\omega)=\sum_{\epsilon,\epsilon^{\prime}}\Pi_{\epsilon}A\Pi_{\epsilon^{\prime}}\delta_{\epsilon-\epsilon^{\prime},\omega}\quad,
\end{equation}
\noindent
and they satisfy 
\begin{equation}\label{eq:7}
    [H,A(\omega)]=-\omega A(\omega), \quad \quad A^{\dagger}(\omega)=A(-\omega).
\end{equation}
\noindent
In terms of these eigenoperators, it can be shown \cite{breuer} that the Lindblad dissipator associated with the microscopic interaction $H_{I}=A\otimes B$ will be, in the rotation wave approximation, 

\begin{equation}\label{eq:8}
    \mathcal{D}(\rho)=\sum_{\omega}\Gamma(\omega)\left[A(\omega)\rho A^{
    \dagger}(\omega)-\frac{1}{2}\{A^{\dagger}(\omega)A(\omega),\rho\}\right].
\end{equation}
\noindent
This method therefore allows to write down the corresponding dissipator. All it requires is sufficient knowledge of the eigenstates of $H$ in order to compute the $A(\omega)$.

Let us now evaluate $\Gamma(\omega)$ in Eq. \eqref{eq:5} for the case of a typical bath interaction operator $B=\sum_{l}g_{l}(a^{\dagger}_{l}+a_{l})$, which appear in Eq \eqref{eq:3}. Using the fact that 
$\left< a^{\dagger}_{l}a_{l'}\right>=\delta_{l,l'}n(\Omega_{l})$, where $n(\alpha)=(e^{\frac{\alpha}{T}}-1)^{-1}$ is the Bose-Einstein distribution. Carrying out the Fourier transform in \eqref{eq:5}, we obtain
\begin{equation*}
\begin{aligned}
    \Gamma&(\omega)=2\pi\sum_{l}g_{l}^{2}\left\{[1+n(\Omega_{l})]\delta(\omega-\Omega_{l})+n(\Omega_{l})\delta(\omega+\Omega_{l})\right\}\\
    =&\int_{0}^{\infty}d\Omega G(\Omega)\left\{[1+n(\Omega_{l})]\delta(\omega-\Omega_{l})+n(\Omega_{l})\delta(\omega+\Omega_{l}) \right\}.
\end{aligned}
\end{equation*}
\noindent
In the last line of the equation above, the sum was transformed into an integral, assuming that the bath frequencies $(\Omega_{l})$ take on a continuum of values. The function $G(\Omega)$ corresponds to $2\pi g_{l}^{2}$ times any aditional factors that come from the transition from a sum to an integral over $\Omega_{l}$ (which do not depend on T). To simplify, we henceforth assume that $G(\Omega)=\gamma$, where $\gamma$ is a constant. We have then

\begin{equation}\label{eq:9}
    \Gamma(\omega)=\begin{cases}
    \gamma[1+n(\omega)], \quad if\quad \omega>0\\
    \gamma n(-\omega), \quad\quad if \quad\omega<0\quad.
    \end{cases}
\end{equation}
A comment is pertinent here. There are other possible spectral densities, for example, the Ohmic case $G(\Omega) = \Omega$. A different density will change the forthcoming computation, but the main result, i.e., the occurrence of rectification shall remain, since, as we see ahead, it is essentially due to the existence of asymmetry and temperature dependent parameters in the system (that change as we invert the baths).

This result is so far general, and valid for any type of bath-coupling operator $A(\omega)$. Now we must specialize it for the case $A=\sigma^{x}_{1}$ and $A=\sigma^{x}_{N}$, which are the coupling operators appearing in Eq.\eqref{eq:3}. This means that we must find the operator $A$ and to do so we need to know the spectral decomposition of $H$.

Now, to diagonalize $H$, we use a fermionic representation through the Jordan-Wigner transformation~\cite{L3,L4} given by:

\begin{equation}\label{eq:10}
    \eta_{l}=Q_{l}\sigma_{l}^{-}\quad,
\end{equation}
\noindent
where the operators $Q_{l}$ are defined by $Q_{l}=\prod_{j=1}^{l-1}(-\sigma_{j}^{z})$.
\noindent
These operators satisfy the fermionic algebra
\begin{equation}\label{eq:11}
    \{\eta^{\dagger}_{l},\eta_{l'}\}=\delta_{l,l'}\quad\quad\{\eta_{l},\eta_{l'}\}=0.
\end{equation}

First of all we transform the Hamiltonian in terms of  $\sigma_{l}^{+}$ and $\sigma_{l}^{-}$ operators given by
\begin{equation}\label{eq:12}
    \begin{aligned}
        \sigma_{l}^{+}&=\frac{1}{2}(\sigma_{l}^{x}+i\sigma_{l}^{y})\\
       \sigma_{l}^{-}&=\frac{1}{2}(\sigma_{l}^{x}-i\sigma_{l}^{y})\quad.\\
    \end{aligned}
\end{equation}

The Hamiltonian in \eqref{eq:1} becomes then:
\begin{equation}\label{eq:13}
    H=\sum_{j=1}^{N}\frac{h_{j}}{2}(\sigma_{j}^{+}\sigma_{j}^{-} -1/2)+\frac{1}{2}\sum_{j=1}^{N-1}\alpha_{j}(\sigma_{j}^{+}\sigma_{j+1}^{-}+\sigma_{j}^{-}\sigma_{j+1}^{+})\quad.
\end{equation}

Using the Jordan-Wigner transformation \eqref{eq:10} we can rewrite \eqref{eq:13} in an quadratic form 

\begin{equation}\label{eq:14}
\begin{aligned}
    H&=\sum_{j=1}^{N}h_{j}\eta_{j}^{\dagger}\eta_{j}+\sum_{j=1}^{N-1}\alpha_{j}(\eta_{j}^{\dagger}\eta_{j+1}+\eta_{j+1}^{\dagger}\eta_{j})\\
    &=\sum_{n,m}W_{n,m}\eta_{n}^{\dagger}\eta_{m}\quad,
    \end{aligned}
\end{equation}
\noindent
where $W_{n,m}$ is a matrix with entries $W_{j,j}=h_{j}$ and $W_{j,j+1}=W_{j+1,j}=\alpha_{j}$.

In order to put $H$ in diagonal form, we first diagonalize the matrix $W$. Since it is symmetric, it may be diagonalized by an orthogonal transformation $S_{n,k}(S^{\dagger}S=1)$ as
\begin{equation}\label{eq:15}
    W_{n,m}=\sum_{k=1}^{N}\epsilon_{k}S_{n,k}S_{m,k}\quad.
\end{equation}
\noindent
the actual form of the eigenvalues and eigenvectors will often be complicate, as they depend on the specific choices of $h_{j}$ and $\alpha_{j}$ in \eqref{eq:1}, which are non-uniform. The eigenvector matrices $S_{n,k}$ will turn out to play an important role as effective coupling constants in the global master equation, see, e.g., 
Eq.\eqref{eq:34}.

Here we define a new set of fermionic operators

\begin{equation}\label{eq:16}
    \Tilde{\eta_{j}}=\sum_{k=1}^{N}S_{j,k}\eta_{k}\quad,
\end{equation}
\noindent
in terms of which the Eq.\eqref{eq:14}  becomes

\begin{equation}\label{eq:17}
    H=\sum_{k=1}^{N}\epsilon_{k}\Tilde{\eta_{k}}^{\dagger}\Tilde{\eta_{k}}\quad.
\end{equation}

Now that we know the diagonal structure of the Hamiltonian, we have to find the operator in the dissipator \eqref{eq:8} in terms of the fermionic operators $A(\omega)$. We start with the left bath, so $A=\sigma_{1}^{x}$. It is easy to see that using \eqref{eq:10} and \eqref{eq:16} we have

\begin{equation}\label{eq:18}
    \sigma_{1}^{x}=\sum_{k=1}^{N}S_{1,k}^{-1}(\Tilde{\eta_{k}}^{\dagger}+\Tilde{\eta_{k}})\quad.
\end{equation}

We note that due to the diagonal structure in Eq \eqref{eq:17}, it follows that $[H,\Tilde{\eta_{k}}]=-\epsilon_{k}\Tilde{\eta_{k}}$. Thus, $\Tilde{\eta_{k}}$ and $\Tilde{\eta_{k}}^{\dagger}$ are eigenoperators of $H$ with allowed transition frequencies $\omega=\epsilon_{k}$ and $\omega=-\epsilon_{k}$ respectively. In this way, we can write the eigenoperator $A(\omega)$

\begin{equation}\label{eq:19}
    A(\omega)=\sum_{k=1}^{N}\left[S_{1,k}^{-1}\left(\Tilde{\eta_{k}}\delta_{\epsilon_{k},\omega}+\Tilde{\eta_{k}}^{\dagger}\delta_{-\epsilon_{k},\omega}\right)\right]\quad.
\end{equation}

The dissipator $\mathcal{D}_{L}(\rho)$, of the left site, is then found from Eq. \eqref{eq:8}

\begin{equation}\label{eq:20}
    \begin{aligned}
        \mathcal{D}_{L}(\rho)&=\sum_{k=1}^{N}\left[\Gamma(\epsilon_{k})(S_{1,k}^{-1})^{2}\left(\Tilde{\eta_{k}}\rho\Tilde{\eta_{k}}^{\dagger}-\frac{1}{2}\{\Tilde{\eta_{k}}^{\dagger}\Tilde{\eta_{k}},\rho\}\right)\right.\\
        \\
        &+\left.\Gamma(-\epsilon_{k})(S_{1,k}^{-1})^{2}\left(\Tilde{\eta_{k}}^{\dagger}\rho\Tilde{\eta_{k}}-\frac{1}{2}\{\Tilde{\eta_{k}}\Tilde{\eta_{k}}^{\dagger},\rho\}\right)\right]\quad.
    \end{aligned}
\end{equation}

Finally, we substitute the expression for $\Gamma$ using Eq. \eqref{eq:9}. In order to do so, we must differentiate the cases where $\epsilon_{k}>0$ and $\epsilon_{k}<0$. We therefore write
\begin{equation}\label{eq:21}
\begin{aligned}
    \mathcal{D}_{L}(\rho)&=\sum_{\epsilon_{k}>0}\gamma (S_{1,k}^{-1})^{2}\left\{[1+n_{L}(\epsilon_{k})]\left[\Tilde{\eta_{k}}\rho\Tilde{\eta_{k}}^{\dagger}-\frac{1}{2}\{\Tilde{\eta_{k}}^{\dagger}\Tilde{\eta_{k}},\rho\right]\right.\\
    &\left.+n_{L}(\epsilon_{k})\left[\Tilde{\eta_{k}}^{\dagger}\rho\Tilde{\eta_{k}}-\frac{1}{2}\{\Tilde{\eta_{k}}\Tilde{\eta_{k}}^{\dagger},\rho\}\right]\right\}\\
    &+\sum_{\epsilon_{k}<0}\gamma (S_{1,k}^{-1})^{2}\left\{n_{L}(-\epsilon_{k})\left[\Tilde{\eta_{k}}\rho\Tilde{\eta_{k}}^{\dagger}-\frac{1}{2}\{\Tilde{\eta_{k}}^{\dagger}\Tilde{\eta_{k}},\rho\}\right]\right. \\
    &\left.+[1+n_{L}(-\epsilon_{k})]\left[\Tilde{\eta_{k}}^{\dagger}\rho\Tilde{\eta_{k}}-\frac{1}{2}\{\Tilde{\eta_{k}}\Tilde{\eta_{k}}^{\dagger},\rho\}\right]\right\},
    \end{aligned}
\end{equation}
where $n_{i}$ is the Bose-Einstein occupation, previously defined.

In Eq.\eqref{eq:21} we see that the separation between positive and negative energies is not good to work with. Instead, we may write the terms in a unified way by defining the Fermi-Dirac occupation
\begin{equation}\label{eq:23}
    f_{i,k}=\frac{1}{e^{\epsilon_{k}}/T_{i}+1}\quad,
\end{equation}
\noindent
and the auxiliary function
\begin{equation}\label{eq:24}
    \chi_{i,k}=2n(|\epsilon_{k}|)+1=\coth\left(\frac{|\epsilon_{k}|}{2T_{i}}\right)\quad,
\end{equation}
\noindent
which we note is always positive. Then the dissipator finally becomes

\begin{equation}\label{eq:25}
\begin{aligned}
    \mathcal{D}_{L}(\rho)&=\sum_{k=1}^{N}\gamma (S_{1,k}^{-1})^{2}\chi_{L,k}\left\{[1-f_{L,k}]\left[\Tilde{\eta_{k}}\rho\Tilde{\eta_{k}}^{\dagger}-\frac{1}{2}\{\Tilde{\eta_{k}}^{\dagger}\Tilde{\eta_{k}},\rho\}\right]\right.\\
    &\left.+f_{L,k}\left[\Tilde{\eta_{k}}^{\dagger}\rho\Tilde{\eta_{k}}-\frac{1}{2}\{\Tilde{\eta_{k}}\Tilde{\eta_{k}}^{\dagger},\rho\}\right]\right\}\quad.
    \end{aligned}
\end{equation}

Now we turn to the bath coupled to the last site $N$. Here the relevant operator is $A=\sigma_{N}^{x}$. In this case, using the Jordan-Wigner transformation and the fact that $\eta_{j}^{\dagger}\eta_{j}=\sigma_{j}^{+}\sigma_{j}^{-}$, we find
\begin{equation}\label{eq:26}
    \sigma_{N}^{x}=Q_{N}(\eta_{N}^{\dagger}+\eta_{N})\quad,
\end{equation}
\noindent
where $Q_{N}=\prod_{j=1}^{N-1}(-\sigma_{j}^{z})$.

Now we define a new operator that counts the total number of fermions 
\begin{equation}\label{eq:27}
    \mathcal{N}=\sum_{i=1}^{N}\eta_{i}^{\dagger}\eta_{i}=\sum_{i=1}^{N}\sigma_{i}^{\dagger}\sigma_{i}=\sum_{i=1}^{N}\Tilde{\eta_{i}}^{\dagger}\Tilde{\eta_{i}}\quad,
\end{equation}
\noindent
here we recall that the number of fermions on the system is proportional to the magnetization in the spin representation. The expression for $\sigma_{N}^{x}$ can be writen as
\begin{equation}\label{eq:26}
    \sigma_{N}^{x}=e^{i\pi\mathcal{N}}e^{-i\pi\eta_{N}^{\dagger}\eta_{N}}(\eta_{N}^{\dagger}+\eta_{N})\quad.
\end{equation}

This expression can be simplified to 
\begin{equation}\label{eq:29}
    \sigma_{N}^{x}=e^{i\pi\mathcal{N}}(\eta_{N}-\eta_{N}^{\dagger})\quad.
\end{equation}
\noindent
As before, using the expression \eqref{eq:16}, we have

\begin{equation}\label{eq:30}
    \sigma_{N}^{x}=\sum_{k=1}^{N}S_{N,k}^{-1}e^{-i\pi\mathcal{N}}(\Tilde{\eta_{k}}-\Tilde{\eta_{k}}^{\dagger})\quad.
\end{equation}

Since $[H,\mathcal{N}]=0$, it follows that $S_{N,k}^{-1}e^{i\pi\mathcal{N}}\Tilde{\eta}_{k}$ is also an eigenoperator with transition frequence $\omega=\epsilon_{k}$, whereas $S_{N,k}^{-1}\Tilde{\eta}_{k}^{\dagger}e^{i\pi\mathcal{N}}$ is an eigenoperator with frequence $\omega=-\epsilon_{k}$. Thus we can write the expression for $A=\sigma_{N}^{x}$
 as
 
 \begin{equation}\label{eq:31}
     A(\omega)=\sum_{k=1}^{N}S_{N,k}^{-1}\left[e^{i\pi\mathcal{N}}\Tilde{\eta}_{k}\delta_{\epsilon_{k},\omega}+\Tilde{\eta}_{k}^{\dagger}e^{i\pi\mathcal{N}}\delta_{-\epsilon_{k},\omega}\right]\quad.
 \end{equation}
 
Following the same previous steps, we can write the dissipator $\mathcal{D}_{R}(\rho)$ as
\begin{equation}\label{eq:32}
\begin{aligned}
    \mathcal{D}_{R}(\rho)=\sum_{k=1}^{N}\gamma &(S_{N,k}^{-1})^{2}\chi_{R,k}\left\{[1-f_{R,k}]\times\right.\\
    &\left[\Tilde{\eta_{k}}e^{i\pi\mathcal{N}}\rho e^{i\pi\mathcal{N}}\Tilde{\eta_{k}}^{\dagger}-\frac{1}{2}\{\Tilde{\eta_{k}}^{\dagger}\Tilde{\eta_{k}},\rho\}\right]\\
    &\left.+f_{R,k}\left[\Tilde{\eta_{k}}^{\dagger}e^{i\pi\mathcal{N}}\rho e^{i\pi\mathcal{N}}\Tilde{\eta_{k}}-\frac{1}{2}\{\Tilde{\eta_{k}}\Tilde{\eta_{k}}^{\dagger},\rho\}\right]\right\}\quad.
    \end{aligned}
\end{equation}

The presence of the global operator $e^{-i\pi\mathcal{N}}$ seems, at a first glance, to complicate matters. However, for all the quantities we shall consider here, and due to the fact that $e^{-2i\pi\mathcal{N}}=1$, that operator will be irrelevant.

Henceforth we define the values
\begin{equation}\label{eq:33}
    g_{L,k}=S_{1,k}^{-1}\quad\quad g_{R,k}=S_{N,k}^{-1}\quad,
\end{equation}
\noindent
for a compact notation.

\section{properties of the steady-state}
\subsection{Occupation Numbers}

With Eq. \eqref{eq:4}, we may now study the behavior of observables such as $\left<\Tilde{\eta_{k}}^{\dagger}\Tilde{\eta_{k'}}\right>$. For the off-diagonal elements $(k\neq k^{\prime})$ we find
\begin{equation*}
    \frac{d}{dt}\left<\Tilde{\eta}_{k}^{\dagger}\Tilde{\eta}_{k'}\right>=-\frac{\gamma}{2}\left(\mathcal{A}_{L,k}+\mathcal{A}_{L,k'}+\mathcal{A}_{R,k}+\mathcal{A}_{R,k'} \right)\left<\Tilde{\eta}_{k}^{\dagger}\Tilde{\eta}_{k'}\right>,
\end{equation*}
\noindent
where $\mathcal{A}_{L(R),k(k')}=g_{L(R),k(k')}\chi_{L(R),k(k')}$.
\newline

Here we see that the term inside parenthesis is always positive, consequently we conclude that $\left<\Tilde{\eta}_{k}^{\dagger}\Tilde{\eta}_{k'}\right>$ will relax exponentially toward zero and therefore vanish at the steady-state. Now for the diagonal elements, again using Eq. \eqref{eq:4}, we find
\begin{equation}\label{eq:34}
    \begin{aligned}
        \frac{d}{dt}\left<\Tilde{\eta}_{k}^{\dagger}\Tilde{\eta}_{k}\right>&=\gamma g_{L,k}\chi_{L,k}(f_{L,k}-\left<\Tilde{\eta}_{k}^{\dagger}\Tilde{\eta}_{k}\right>)\\
        &+\gamma g_{R,k}\chi_{R,k}(f_{R,k}-\left<\Tilde{\eta}_{k}^{\dagger}\Tilde{\eta}_{k}\right>)\quad.
    \end{aligned}
\end{equation}

With Eq. \eqref{eq:34} we can see that, in the steady-state, the occupations will converge to
\begin{equation}\label{eq:35}
    \left<\Tilde{\eta}_{k}^{\dagger}\Tilde{\eta}_{k}\right>=\frac{g_{L,k}\chi_{L,k}f_{L,k}+g_{R,k}\chi_{R,k}f_{R,k}}{g_{L,k}\chi_{L,k}+g_{R,k}\chi_{R,k}}\quad.
\end{equation}
\noindent
When $T_{L}=T_{R}$ this reduces to $\left<\Tilde{\eta}_{k}^{\dagger}\Tilde{\eta}_{k}\right>=f_{k}$ as expected. Now let us see what happens if the chain is subjected to a small difference of temperature, given by $T_{L}=T+\delta T/2$ and $T_{R}=T-\delta T/2$. Equation \eqref{eq:35} reduces to
\begin{equation}
    \left<\Tilde{\eta}_{k}^{\dagger}\Tilde{\eta}_{k}\right>\simeq f_{k}+\frac{\delta T}{2}\left(\frac{g_{L,k}-g_{R,k}}{g_{L,k}+g_{R,k}}\right)\frac{\partial f_{k}}{\partial T} + \mathcal{O}(\delta T)^{2}.
\end{equation}
\noindent

If the chain is homogeneous then, by symmetry, $g_{L,k}=g_{R,k}$ and the first correction will be of order $\delta T^{2}$. This is expected since, for a homogeneous chain, the perturbation should not depend on the sign of $\delta T$. But we see that, in general, when we have a inhomogeneous chain, reversing the order of the baths will change the occupation numbers.

With Eq.\eqref{eq:24} we can analyze the behavior of the occupations numbers. We can see that the relaxation in Eq.\eqref{eq:34} will occur with typical rates proportional to $\chi_{i,k}$. We note that this function diverges when the energy approaches zero. Thus, the present model predicts that different modes of the Hamiltonian will relax with different rates, the relaxation being faster the smaller is the energy of the fermionic mode. This fact is actually quite reasonable from a physical standpoint. The energy $\epsilon_{k}$ of a fermionic mode represents the energy gap that needs to be overcomed in a thermal transition. Modes with small gap should experience a larger number of transitions while they relax to equilibrium and therefore should relax more quickly.

\subsection{Particle and Energy Current}

Using Eq.\eqref{eq:4} we can derive some expressions for the energy and particle currents. In the fermionic representation, the temperature unbalance between the two baths will lead to a flow of particles along the chain. In the spin representation, this is mapped into a flow of magnetization.

To evaluate the current of particles/magnetization, we start with a conservation law for the time evolution of $\left<\mathcal{N}\right>$. Since $[H,\mathcal{N}]=0$, it follows from Eq.\eqref{eq:4} that

\begin{equation}\label{eq:37}
    \frac{d}{dt}\left<\mathcal{N}\right>=tr \left\{ \mathcal{N}\mathcal{D}_{L}(\rho)\right\}+tr \left\{ \mathcal{N}\mathcal{D}_{R}(\rho)\right\}.
\end{equation}
\noindent
The two terms on the right-hand side may be readily identified as the flow of particles from the system to each of the reservoirs. In the steady state we have $d\left<\mathcal{N}\right>/dt=0$ and both fluxes will coincide. We then define

\begin{equation}\label{eq:38}
    J_{\mathcal{N}}=tr\left\{ \mathcal{N}\mathcal{D}_{L}(\rho)\right\}=-tr \left\{ \mathcal{N}\mathcal{D}_{R}(\rho)\right\}\quad,
    \end{equation}
which is, we stress, a relation valid in the steady state.

Using Eq.\eqref{eq:25} for the dissipator, we find that
\begin{equation}\label{eq:39}
    J_{\mathcal{N}}=\sum_{k=1}^{N}\gamma g_{L,k}\chi_{L,k}\left[f_{L,k}-\left<\Tilde{\eta}_{k}^{\dagger}\Tilde{\eta}_{k'}\right>\right]\quad.
\end{equation}
Substituting the occupation for the steady-state we have
\begin{equation}\label{eq:40}
    J_{\mathcal{N}}=\sum_{k=1}^{N}\gamma\frac{g_{L,k}\chi_{L,k}g_{R,k}\chi_{R,k}}{g_{L,k}\chi_{L,k}+g_{R,k}\chi_{R,k}}(f_{L,k}-f_{R,k})\quad.
\end{equation}

We see that the current is essentially a sum of all occupations unballances, weighted by certain functions. It is important to note that these weights are temperature dependent. Precisely, we see that the current is nothing but a sum of currents associated to each eigenmode of the system.

Now we can define the energy current doing the same steps in terms of the conservation of $\left<H\right>$. Its form will be analogous to Eq.\eqref{eq:39} and \eqref{eq:40}, but each term now will be multiplied by $\epsilon_{k}$:

\begin{equation}\label{eq:41}
    J_{E}=\sum_{k=1}^{N}\gamma\epsilon_{k}\frac{g_{L,k}\chi_{L,k}g_{R,k}\chi_{R,k}}{g_{L,k}\chi_{L,k}+g_{R,k}\chi_{R,k}}(f_{L,k}-f_{R,k})\quad.
\end{equation}

With the expression for the energy current, we can analyze the occurrence of rectification on the system. First of all let us analyze the expression for the particle current given by Eq.\eqref{eq:40}. For a small temperature gradient, it becomes 
\begin{equation}\label{eq:42}
    \begin{aligned}
        J_{\mathcal{N}}&\simeq\gamma\delta T\sum_{k=1}^{N}\frac{g_{L,k}g_{R,k}}{g_{L,k}+g_{R,k}}\chi_{k}\frac{\partial f_{k}}{\partial T}\\
        &+\gamma \delta T^{2}\sum_{k=1}^{N}\frac{g_{L,k}g_{R,k}(g_{R,k}-g_{L,k})}{(g_{L,k}+g_{R,k})^{2}}\frac{\partial\chi_{k}}{\partial T}\frac{\partial f_{k}}{\partial T}\quad.
    \end{aligned}
\end{equation}
\noindent
When a system does not present rectification, the current will be an odd function of $\delta T$. Here we see the presence of a term proportional to $\delta T^{2}$, which will be the lowest order contribution to the rectification. Note that it will be non-zero when $g_{L,k}\neq g_{R,k}$.

We can write down these result more explicitly, using \eqref{eq:23} and \eqref{eq:24}. Finally, we define:

\begin{equation}\label{eq:43}
    J_{\mathcal{N}}=\gamma\delta T J_{1}+\gamma \delta T^{2} J_{2}+...\quad,
\end{equation}
\noindent
where
\begin{equation}\label{eq:44}
    J_{1}=\sum_{k=1}^{N}\frac{g_{L,k}g_{R,k}}{g_{L,k}+g_{R,k}}\frac{|\epsilon_{k}|}{2T^{2}}\csch{\left(\frac{\epsilon_{k}}{T}\right)}
\end{equation}
and
\begin{equation}\label{eq:45}
    J_{2}=\sum_{k=1}^{N}\frac{g_{L,k}g_{R,k}(g_{R,k}-g_{L,k})}{(g_{L,k}+g_{R,k})^{2}}\frac{\epsilon_{k}|\epsilon_{k}|}{2T^{4}}\csch^{2}\left(\frac{\epsilon_{k}}{T}\right)\quad.
\end{equation}

Here we note that $J_{2}$ is the remaining term in $\mathcal{O}(\Delta T)$ for the occurrence of thermal rectification. As we have a inhomogeneous chain, $(g_{L,k}-g_{R,k})\neq 0$.

\section{Heat Rectification}

With the expressions for the energy current, we can investigate the occurrence of thermal rectification in the $XX$ chain. We know from the first law of thermodynamics that energy current is given by the power current and the heat current:

\begin{equation}\label{eq:46}
    \Dot{E}=\Dot{W}+\sum_{r}\Dot{Q}_{r}\quad,
\end{equation}
\noindent
where $r=L,R$ represents the index of the baths.

From the microscopic derivation for the Lindblad equation we can calculate these quantities: 
\begin{equation}\label{eq:47}
    \begin{aligned}
            \Dot{W}(t)&=Tr\left\{\Dot{H}_{S}(t)\rho_{S}\right\}\\
            \Dot{Q}_{r}(t)&=Tr\left\{H_{S}(t)\mathcal{D}_{r}(\rho_{S})\right\}\quad.
    \end{aligned}
\end{equation}

We can see that our Hamiltonian is independent of time, so no work can be done on the system and the energy current is given by heat current
\begin{equation}\label{eq:48}
    \Dot{E}=\sum_{r}\Dot{Q}_{r}\equiv\Dot{Q}\quad.
\end{equation}
\noindent
These definitions are justified, for example, in \cite{barra}.

According to Eq.\eqref{eq:41} we have to calculate the eigenvalues and eigenvectors of the matrix associated to the Hamiltonian to compute the heat current. That is, we have to diagonalize a inhomogeneous tridiagonal matrix. The need to introduce more complex asymmetries and interactions makes any analytical treatment for this problem much more difficult. 

In order to simplify the interaction matrix and to find an analytical solution, we consider a system subject to a perturbation on the external magnetic field in the first and last sites. The matrix $W$ describing the interaction is given by
\begin{equation}\label{eq:49}
W=\begin{pmatrix}
h - \alpha&\alpha& 0& \ldots & 0& \ldots & 0\\
\alpha&h&\alpha& \ddots & 0 & \ldots& 0\\
0&\alpha&h& \alpha & 0 & \ldots& 0\\
\vdots & \ddots & \ddots & \ddots & \ddots & \ddots & \vdots\\
0 & 0 &0 &\alpha& h &\alpha& 0\\
\vdots & \vdots & \vdots & \ddots & \alpha & \ddots & \alpha\\
0 & 0 & 0 & \ldots & 0& \alpha& h+\alpha\\
\end{pmatrix}\quad .
\end{equation}
\noindent
To be clear: the external magnetic field is given by $h$ for the internal sites $i\in[2,N-1]$. Here,  $\alpha$ represents the interaction between the neighbors that we assume to be constant, and in the first and last sites has a perturbation given by $\alpha$, the same value of the interaction between the sites.

For this specific matrix we have an analytical solution for the eigenvalues and eigenvectors: 

\begin{equation}\label{eq:50}
\begin{aligned}
    \epsilon_{k}&=h+2\alpha \cos\left[\frac{(2k-1)\pi}{2N}\right]\\
    v_{j}^{k}&=\sin\left[\frac{(2j-1)(2k-1)\pi}{4N}\right]
    \end{aligned}
\end{equation}
\noindent
where $k,j=1, 2,...,N$.
All the details and the process of diagonalization can be found in \cite{willsol}.

To proceed with the calculation, first of all we need to normalize the eigenvector. By using the geometric sum we obtain
\begin{equation}\label{eq:51}
    v_{j}^{k}=\sqrt{\frac{2}{N}}\sin\left[\frac{(2j-1)(2k-1)\pi}{4N}\right]\quad.
\end{equation}
Thus, the matrix $S$ that diagonalizes the matrix $W$, Eq.\eqref{eq:15}, is given by
\begin{equation}\label{eq:52}
    S_{j,k}=\sqrt{\frac{2}{N}}\sin\left[\frac{(2j-1)(2k-1)\pi}{4N}\right]\quad.
\end{equation}

Since $S$ is orthogonal, we can calculate the quantities given by \eqref{eq:33}. After several manipulations we have
\begin{equation}\label{eq:53}
\begin{aligned}
    g_{L,k}&=\frac{2}{N}\sin^{2}\left[\frac{(2k-1)\pi}{4N}\right]\\
    \\
    g_{R,k}&=\frac{2}{N}\cos^{2}\left[\frac{(2k-1)\pi}{4N}\right]\quad.
\end{aligned}
\end{equation}

Substituting \eqref{eq:51} and \eqref{eq:53} in the expression for the heat flux \eqref{eq:41} we arrive at
\begin{equation}\label{eq:54}
\begin{aligned}
    J=\frac{2\gamma}{N}\sum_{k=1}^{N}&\left[\frac{\sin^{2}(\delta_{k})\cos^{2}(\delta_{k})\chi_{L,k}\chi_{R,k}}{\cos^{2}(\delta_{k})\chi_{L,k}+\sin^{2}(\delta_{k})\chi_{R,k}}\right]\times\\
    \\
    &\times\left[h+2\alpha \cos(2\delta_{k})\right](f_{L,k}-f_{R,k})
    \end{aligned}\quad,
\end{equation}
\noindent
where $\delta_{k}=\frac{(2k-1)\pi}{4N}$ and $J_{E}\equiv J$.
\newline

To investigate the occurrence of rectification we have to compute the heat flux in the reversed bias. To compute these values, we change the baths. This represents the exchange of the temperatures $T_{L}^{\prime}=T_{R}$ and $T_{R}^{\prime}=T_{L}$. According to \eqref{eq:41} and \eqref{eq:54}, the heat flow in the opposite direction is
\begin{equation}\label{eq:55}
\begin{aligned}
    J_{r}=-\frac{2\gamma}{N}\sum_{k=1}^{N}&\left[\frac{\sin^{2}(\delta_{k})\cos^{2}(\delta_{k})\chi_{L,k}\chi_{R,k}}{\cos^{2}(\delta_{k})\chi_{R,k}+\sin^{2}(\delta_{k})\chi_{L,k}}\right]\times\\
    \\
    &\times\left[h+2\alpha \cos(2\delta_{k})\right](f_{L,k}-f_{R,k})
    \end{aligned},
\end{equation}
\noindent
where the index $r$ means the reversed flow.

As we can see, the expressions \eqref{eq:54} and \eqref{eq:55} have a complex dependence on the temperature, given by $\chi_{L(R),k}$, and the analytical treatment from these expressions is a complicated task. In order to simplify the analytical calculations, we make some additional assumptions.

As we can see in the expression \eqref{eq:50} for the eigenvalues, we can have a spectrum that is entirely positive by taking $h>0$, $\alpha>0$ and $h>2\alpha$. Also, regarding the baths, we take our system subjected to a large temperature gradient. Namely, we consider the limits $T_{L}\rightarrow \infty $ and  $T_{R}\rightarrow 0$.

From these assumptions, we have to analyze the behavior of \eqref{eq:23} and \eqref{eq:24} to calculate the heat flux. According to the Fermi-Dirac occupation, we can see that, when $T_{L}\rightarrow \infty $ and  $T_{R}\rightarrow 0$,
\begin{equation}\label{eq:56}
    \begin{aligned}
        &f_{L,k}\rightarrow 1/2\\
        &f_{R,k}\rightarrow 0\quad.\\
    \end{aligned}
\end{equation}
Carrying out the same analysis for $\chi_{L,k}$ and $\chi_{R,k}$ given by \eqref{eq:24}, we have
\begin{equation}\label{eq:57}
    \begin{aligned}
        &\chi_{L,k}\rightarrow \infty\\
        &\chi_{R,k}\rightarrow 1\quad.\\
    \end{aligned}
\end{equation}
Replacing these results in the expression for the heat flow \eqref{eq:54} and the reversed heat flow \eqref{eq:55}, we obtain
\begin{equation}\label{eq:58}
    \begin{aligned}
        J&=\frac{\gamma}{N}\sum_{k=1}^{N}\left( h+2\alpha \cos\left[\frac{(2k-1)\pi}{2N}\right]\right)\cos^{2}\left[\frac{(2k-1)\pi}{4N}\right]\\
        J_{r}&=-\frac{\gamma}{N}\sum_{k=1}^{N}\left( h+2\alpha \cos\left[\frac{(2k-1)\pi}{2N}\right]\right)\sin^{2}\left[\frac{(2k-1)\pi}{4N}\right].
    \end{aligned}
\end{equation}
These fluxes can be calculated by using the geometric sum. After some algebraic manipulations we find
\begin{equation}\label{eq:59}
    \begin{aligned}
        J=&\frac{\gamma (h+\alpha)}{2}\\
        J_{r}=&-\frac{ \gamma(h-\alpha)}{2}\quad.\
    \end{aligned}
\end{equation}
\noindent
with $\alpha\neq0$ \cite{willsol, obs}.

As the values are different ($J\neq J_{r}$), we have the existence of thermal rectification. By \eqref{eq:59} we can see that we have a ballistic transport, that is, the heat flow does not depend on the size of the chain. It is interesting to note that, in such a regime, the difference between the magnitude of the flows depends only on $\alpha$. Consequently, we note an important result: the rectification factor remains finite when $N\rightarrow\infty$.

We can perform the same analysis for a negative spectrum. Now we consider $h<0$, $\alpha>0$ and $|h|>2\alpha$. The procedure is the same as we  previously described,
\begin{equation}\label{eq:60}
    \begin{aligned}
        J&=-\frac{\gamma}{N}\sum_{k=1}^{N}\left( h+2\alpha \cos\left[\frac{(2k-1)\pi}{2N}\right]\right)\left[\cos^{2}\left[\frac{(2k-1)\pi}{4N}\right]\right]\\
        J_{r}&=\frac{\gamma}{N}\sum_{k=1}^{N}\left( h+2\alpha \cos\left[\frac{(2k-1)\pi}{2N}\right]\right)\left[\sin^{2}\left[\frac{(2k-1)\pi}{4N}\right]\right].
    \end{aligned}
\end{equation}

The heat flows are given by
\begin{equation}\label{eq:61}
    \begin{aligned}
        J=&\frac{\gamma (|h|-\alpha)}{2}\\
        J_{r}=&-\frac{ \gamma(|h|+\alpha)}{2}\quad.\
    \end{aligned}
\end{equation}

As expected, we have thermal rectification. 
\newline

In more asymmetric systems we expect the improvement of the rectifcation. 

Now we analyze the regime of strong interaction between the sites. If we take $\alpha$ large enough, we split the energy spectrum in positive and negative values. More specifically, for $N$ even, if we assume
\begin{equation*}
    \alpha>\frac{h}{2}\left |\sec\left[\frac{(N+1)\pi}{2N}\right]\right |\quad,
\end{equation*}
\noindent
the spectrum is divided into $N/2$ positive values and $N/2$ negative ones. In the regime of large temperature gradient in \eqref{eq:56} and \eqref{eq:57}, we have the following expression for the heat current

\begin{equation}\label{eq:62}
    J=\frac{\gamma}{2}\left[\sum_{k=1}^{N/2}\epsilon_{k}g_{R,k}-\sum_{k=N/2+1}^{N}\epsilon_{k}g_{R,k}\right]\quad,
\end{equation}
\noindent
where $g_{R,k}$ is given by \eqref{eq:53}. Carrying out the manipulations, we find the following heat current:
\begin{equation}\label{eq:63}
    J=\frac{\gamma}{N}\csc\left(\frac{\pi}{2N}\right)\left[\alpha+ \frac{h}{2}\right]\quad.
\end{equation}

For the heat current in the reversed bias, we have the following expression
\begin{equation}\label{eq:64}
    J_{r}=-\frac{\gamma}{2}\left[\sum_{k=1}^{N/2}\epsilon_{k}g_{L,k}-\sum_{k=N/2+1}^{N}\epsilon_{k}g_{L,k}\right]\quad.
\end{equation}

Performing the algebric manipulations, we find
\begin{equation}\label{eq:65}
    J_{r}=-\frac{\gamma}{N}\csc\left(\frac{\pi}{2N}\right)\left[\alpha-\frac{h}{2} \right]\quad.
\end{equation}

Again, we have thermal rectification, and comparing with the result obtained in \eqref{eq:59}, we see that, for strong interaction $\alpha$, the difference between the magnitude of the flows depends now on the magnetic field $h$. Moreover, note that, again, we have balistic transport of heat, since as $N\rightarrow\infty$ the currents converge to non-zero values ($\frac{\gamma}{N}csc(\frac{\pi}{2N})\rightarrow\frac{2\gamma}{\pi})$, and again we have a finite rectification factor in the thermodynamic limit.

 If we define the following rectification factor:
\begin{equation}\label{eq:66}
    \mathcal{R}=\frac{J+J_{r}}{\min(J,|J_{r}|)}\quad,
\end{equation}
\noindent
we can write for \eqref{eq:59}:

\begin{equation*}
    \mathcal{R}=\frac{2\alpha}{h-\alpha}
\end{equation*}
\noindent
and for \eqref{eq:63} with \eqref{eq:65}

\begin{equation*}
    \mathcal{R}=\frac{h}{\alpha-h/2}\quad.
\end{equation*}

In the next section we perform some numerical analysis to investigate the behavior of rectification in some interesting and more intricate cases using \eqref{eq:66} for the rectification factor.

\section{Numerical analysis}

In this section we perform some numerical analysis using the expressions for the heat flow given by \eqref{eq:41}. We compute the exact eigenvalues and the eigenvectors for an inhomogeneous matrix that represents the interaction of our system \eqref{eq:14}. We investigate different systems, for example, models given by the sequential coupling of
parts with different interactions as the usual proposal of thermal diodes
\cite{BLiRMP, Terraneo, LiC}, or graded systems \cite{WPC, SPL}, which are other recurrent models in this field.  

\begin{figure}[!htbp]
    \centering
    \includegraphics[width=\columnwidth]{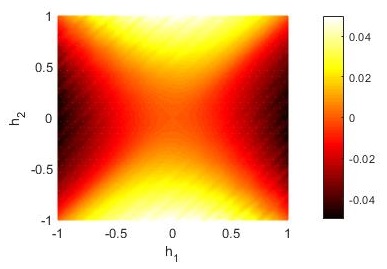}
    \caption{Rectification profile for a junction of external magnetic fields \eqref{eq:67} composed by 50 sites. The difference of temperature is given by $\Delta T=T_{L}-T_{R}=5$ while the interaction is $\alpha_{i}=1$.}
    \label{FIG1}
\end{figure}

We perform the first analysis by varying the external magnetic field and keeping fixed the interaction between the sites of the chain ($\alpha_{i}=1$). We consider a system subjected to two different external magnetic fields:
\begin{equation}\label{eq:67}
\begin{aligned}
    &h_{i}=h_{1}\quad\quad i\in[1,...,N/2]~,\\
    &h_{i}=h_{2}\quad\quad i\in[N/2+1,...,N]~.
    \end{aligned}
\end{equation}
The rectification profile for a system of 50 sites is depicted in Fig.\ref{FIG1}.

\begin{figure}[!htbp]
    \centering
    \includegraphics[width=\columnwidth]{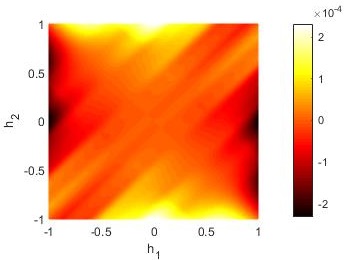}
    \caption{Rectification profile for a junction of external magnetic fields \eqref{eq:67} composed by 50 sites. The difference of temperature is given by $\Delta T=T_{L}-T_{R}=5$ while the interaction is $\alpha_{i}=5$.}
    \label{FIG2}
\end{figure}

If we make the interaction between the sites more intense, we see more nuances in the rectification profile and also a decrease in rectification intensity, as presented in Fig.\ref{FIG2}.

We also study the behavior of the rectification with the interaction between the sites $(\alpha_{i})$. First of all, we investigate the existence of rectification without external magnetic field, $h_{i}=0$. We  consider a system composed by 50 sites subjected to a difference of temperature $\Delta T=T_{L}-T_{R}=5$. 

Here we  consider a system composed by two different values of interactions: 
\begin{equation}\label{eq:68}
\begin{aligned}
    &\alpha_{i}=\alpha_{1}\quad\quad i\in[1,...,N/2]~,\\
    &\alpha_{i}=\alpha_{2}\quad\quad i\in[N/2+1,...,N]~.
    \end{aligned}
\end{equation}

\begin{figure}[!htbp]
    \centering
    \includegraphics[width=\columnwidth]{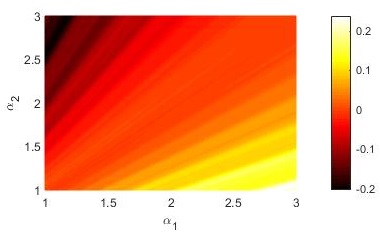}
    \caption{Rectification profile for a junction of two interactions \eqref{eq:68}. The system is composed by 50 sites and the difference of temperature is given by $\Delta T=T_{L}-T_{R}=5$, without magnetic field.}
    \label{FIG3}
\end{figure}

The result for the system in Eq.\eqref{eq:68} is given by Fig.\ref{FIG3}.

As we can see in Fig.\ref{FIG3}, we have rectification in a system only changing the interaction $\alpha_{i}$. i.e, the existence of an external magnetic field is not essential for the occurrence of thermal rectification.

\begin{figure}[!htbp]
    \centering
    \includegraphics[width=\columnwidth]{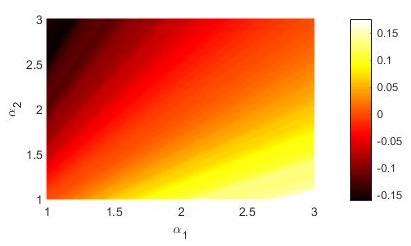}
    \caption{Rectification profile for a junction of two interactions \eqref{eq:68}. The system is composed by 50 sites and the difference of temperature is given by $\Delta T=T_{L}-T_{R}=5$. The magnetic field is fixed at $h_{i}=5$.}
    \label{FIG4}
\end{figure}

Now we investigate the behavior with the external magnetic field. We perform the same calculations with a constant external magnetic field, fixed at $h_{i}=5$. The result is presented in Fig.\ref{FIG4}.

We see that the rectification is more sensitive to changes in the external magnetic field compared to changes in the interaction between neighbor sites. We can observe in these rectification profiles that we have a reversal of rectification, that is, there are values of $h_{i}$ and $\alpha_{i}$ such that the rectification value changes sign. This phenomenon is discussed in Refs. \cite{reversal, EP2}.

Another common way to construct a thermal diode is the use of graded materials. These materials are abundant in nature and can be manufactured. Hence, we investigate the behavior of thermal rectification in chains with graded structure, i.e, a system in which its internal parameters gradually varies in space.

\begin{figure}[!htbp]
    \centering
    \includegraphics[width=\columnwidth]{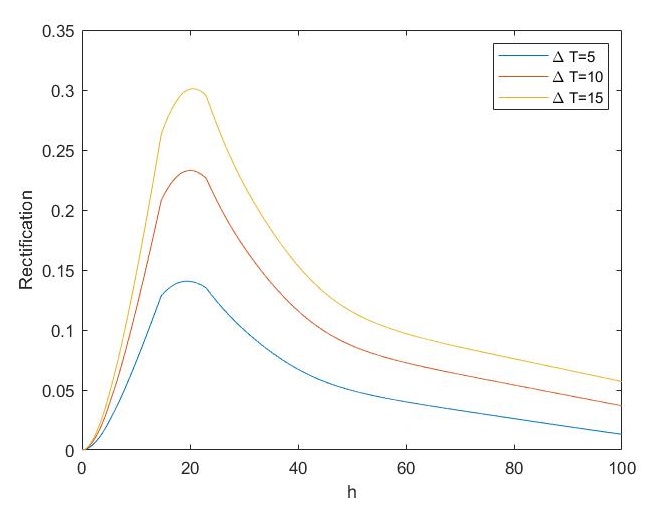}
    \caption{Rectification profile for a linear graded external magnetic field. The system is composed by 10 sites and the difference of temperature is given by $\Delta T=5,10$ and $15$. The intersite interaction is fixed at $\alpha_{i}=1$. The external graded magnetic field is given by $h_{i}\varpropto ih$.}
    \label{FIG5}
\end{figure}

For a graded external magnetic field, we have the pattern for a system composed by 10 sites as presented in Fig.\ref{FIG5}.

\begin{figure}[!htbp]
    \centering
    \includegraphics[width=\columnwidth]{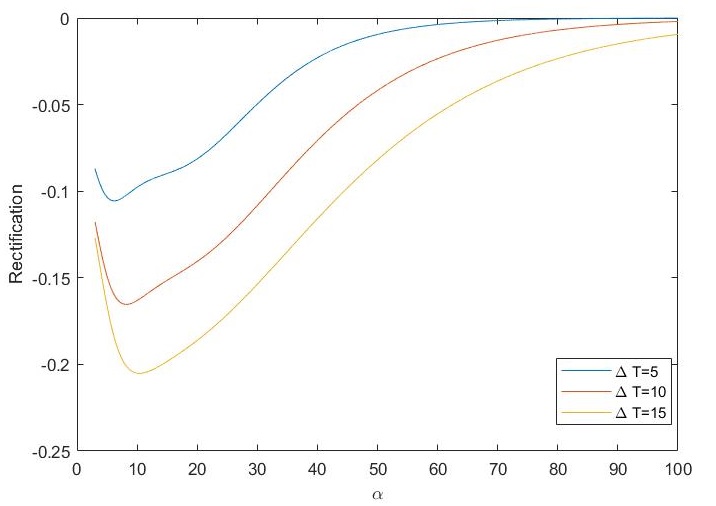}
    \caption{Rectification profile for a linear graded interaction. The system is composed by 10 sites and the difference of temperature is given by $\Delta T=5,10$ and $15$. The external magnetic field is fixed at $h_{i}=5$ and the intersite interaction is given by $\alpha_{i}\varpropto i\alpha$.}
    \label{FIG6}
\end{figure}

The rectification profile for a graded intersite interaction is depicted in Fig.\ref{FIG6}.

\begin{figure}[!htbp]
    \centering
    \includegraphics[width=\columnwidth]{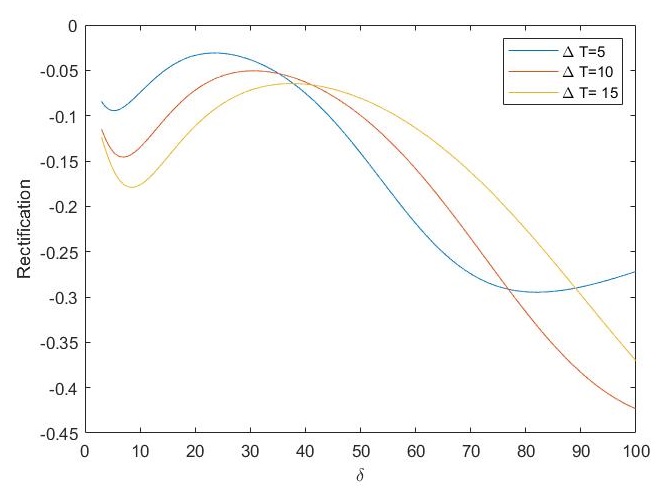}
    \caption{Rectification profile for a linear graded chain. The external magnetic field and the inter site interaction are linear on $\delta$, more specifically, $h_{i}\varpropto i\delta$ and $\alpha_{i}\varpropto i\delta$. The system is composed by 10 sites and the difference of temperature is given by $\Delta T=5,10$ and $15$.}
    \label{FIG7}
\end{figure}

Now, if we make a graded external magnetic field and graded inter site interaction, we find the profile depicted in Fig.\ref{FIG7}. 

In conclusion, we see that the rectification is more significant if we make the temperature gradient more intense, Fig.\ref{FIG5} and Fig.\ref{FIG6}. In Fig.\ref{FIG7} we see that a graded structure changes all the pattern of rectification as well as its intensity, compared to Fig.\ref{FIG5} and Fig.\ref{FIG6}.

\section{Final Remarks}

In the present paper, aiming to understand the mechanism of thermal rectification in quantum systems, we investigated in detail the heat current
in the $XX$ chain with nearest neighbor interactions and global dissipators,
a simple quadratic spin model. We showed the existence of thermal rectification even for a simple case of a slightly asymmetrical chain. Interestingly, we give examples of rectification that remains finite as 
the system length increases, i.e., it does not vanish in the limit $N \rightarrow \infty$.

In relation to the possible experimental realization of such models, we 
recall the possibility to engineer $XXZ$ chains with different configurations, i.e., with different  values for the coefficientes of $\sigma_{j}^{x}\sigma_{j+1}^{x}, \sigma_{j}^{y}\sigma_{j+1}^{y}$ and 
$\sigma_{j}^{z}\sigma_{j+1}^{z}$ \cite{endres, barredo}. We also recall
the simulation of these Heisenberg models by means of cold atoms in optical
lattices \cite{bloch} or trapped ions \cite{blatt}. And experiments with
Rydberg atoms in optical traps involving these spin models are presented 
in Ref.\cite{duan, whitlock, PhysRevX}.

A further comment is pertinent. For other types of dissipators, e.g., for
those local dissipators that target polarization at the boundaries of the chain, the energy current is not only heat as it happens here, but it 
consists of heat and work (power). Such a distinction is crucial for thermodynamic consistency. A detailed discussion is presented in 
Refs.\cite{FBarra, Pereira2018, GL-NJP}.

To conclude, with the results presented here we believe to shed some light
in the problem of quantum thermal diodes proposals: the occurrence of a
robust thermal rectification in this simple model shows that rectification in quantum spin systems is an ubiquitous phenomenon. 

\vspace{1cm}

{\bf Acknowledgment:} Work partially supported by CNPq (Brazil).


\newpage

\begin{thebibliography}{100}
	
	\bibitem{BLiRMP}
	N. Li, J. Ren, L. Wang, G. Zhang, P. H\"{a}nggi, and B. Li,
	Rev. Mod. Phys. \textbf{84}, 1045 (2012).
	
	\bibitem{Deb} P. Debye, ``Vortraege ueber die Kinetsche Theorie der Materie und der Elektrizitaet'' (Leipzig: Teubner, 1914).
	
	\bibitem{Pei} R. Peierls, Ann. Physik \textbf{3},  1055 (1929).
	
	\bibitem{BLL} F. Bonetto, J. L. Lebowitz, and J. Lukkarinen, J. Stat. Phys. \textbf{116}, 783 (2004).
	
	\bibitem{PLA} E. Pereira, H. C. F. Lemos, and R. R. \'Avila, Phys. Rev. E \textbf{84}, 061135 (2011).
	
	
	\bibitem{RLL} Z. Rieder, J. L. Lebowitz, and E. Lieb, J. Math. Phys. 
	\textbf{8}, 1073 (1967).
	
    \bibitem{WPC} J. Wang, E. Pereira, and G. Casati, Phys. Rev. E 
    \textbf{86}, 010101(R) (2012).


    \bibitem{GL1} G. T. Landi, E. Novais, M. J. de Oliveira, and D. Karevski, Phys. Rev. E \textbf{90}, 042142 (2014).
    
        \bibitem{SPL} L. Schuab, E. Pereira, and G. T. Landi, Phys. Rev. E 
     \textbf{94}, 042122 (2016).
     
       \bibitem{P1} E. Pereira, Phys. Rev. E \textbf{99}, 032116 (2019).
    
    \bibitem{breuer} H.-P. Breuer and F. Petruccione, Albert-Ludwigs-Universitdt Freiburg, Fakultdt fiir Physik and Istituto Italiano per gli Studi Filosofici, The Theory of Open Quantum Systems, Oxford University Press.
    
    \bibitem{L3} E. Lieb, T. Schultz, and D. Mattis, Annals of Physics \textbf{466}, 407 (1961).
    \bibitem{L4} E. Lieb, T. Schultz, and D. Mattis, Reviews of Modern Physics \textbf{36}, 856 (1964).


     
     \bibitem{barra} F. Barra, Scientific Reports \textbf{5}, 20452322 (2015)
     
     \bibitem{willsol} A. R. Willms, Journal on Matrix Analysis and Applications \textbf{30} 639-656 (2008)
     
     \bibitem{obs} As stressed, we cannot take $\alpha=0$ in the expressions for the rectification (derived for nonzero $\alpha$). In the case of $\alpha=0$, we have a homogeneous magnetic field and no interactions between the sites, and so we need to go back to 
     Eq.\eqref{eq:41} for the energy flux. For this case, the eigenvectors of the matrix interaction W are equal to the canonical eigenvectors, and so $g_{L,k}$ and $g_{R,k}$ are equal to 0 or 1, depending on $k$. Analyzing the Eq.\eqref{eq:40}, with this behavior in $k$, we find $J_{E}=0$, as expected.
     
     	\bibitem{Terraneo} M. Terraneo, M. Peyrard, and G. Casati, Phys. Rev. Lett. {\bf 88}, 094302 (2002).
	
	\bibitem{LiC} B. Li, L. Wang, G. Casati, Phys. Rev. Lett. \textbf{93}, 184301 (2004).

	\bibitem{reversal} Zhang, Lifa and Yan, Yonghong and Wu, Chang Qin and Wang, Jian Sheng and Li, Baowen, Phys. Rev. B \textbf{80} 10980121 (2009)
	
	\bibitem{EP2} A. L. de Paula, E. Pereira, R. C. Drumond and M. C. O. Aguiar, J. Phys. Cond. Mat. \textbf{32}, 175403 (2020).
	
	

	
	\bibitem{endres} M. Endres, H. Bernien, A. Keesling, H. Levine, E. R. Anschuetz, A. Krajenbrink, C. Senko, V. Vuletic, M. Greiner, G. Markus and M. D. Lukin, Science \textbf{354}, aah3752 (2016).
	
	\bibitem{barredo} D. Barredo, S. De L{\'e}s{\'e}leuc, V. Lienhard, T. Lahaye, and A. Browaeys,  Science \textbf{354}, aah3778 (2016).
	
	\bibitem{bloch} I. Bloch, J. Dalibard, and S. Nascimbene,  Nat. Phys. \textbf{8}, 267 (2012).
	
	\bibitem{blatt} R. Blatt and  C. F. Roos, Nat. Phys. \textbf{8}, 277 (2012).
	
	\bibitem{duan} L-M. Duan, E. Demler, and M. D. Lukin, Phys. Rev. Lett. \textbf{91}, 090402 (2003).
	
	
	
	\bibitem{whitlock} S. Whitlock, A. W. Glaetzle,  and P. Hannaford, J. Phys. B \textbf{50}, 074001 (2017).
	
	\bibitem{PhysRevX} T. L. Nguyen, J. M.  Raimond, C. Sayrin, R. Corti\~nas, T. Cantat-Moltrecht, F. Assemat, I. Dotsenko, S. Gleyzes, S.  Haroche, G. Roux, Th. Jolicoeur,  and M. Brune, Phys. Rev. X \textbf{8},
	011032 (2018).
	

\bibitem{FBarra} F. Barra, Sci. Rep. {\bf 5}, 14873 (2015).	

\bibitem{Pereira2018} E. Pereira, Phys. Rev. E {\bf 97}, 022115 (2018).

\bibitem{GL-NJP} G. De Chiara {\it et al.}, New J. Phys. {\bf 20}, 113024 (2018).	
	
	
	
	
	
	
	
	
	
	
	
	
	
	
	
	

	
	
	
	
	
	
\end{thebibliography}
\end{document}